\documentclass[%
 reprint,
superscriptaddress,
 amsmath,amssymb,
 aps,
]{revtex4-1}
\usepackage[latin1]{inputenc}
\usepackage{graphicx}
\usepackage{dcolumn}
\usepackage{bm}
\usepackage{siunitx}
\usepackage{hyperref}
\newcommand{\picpath}{.}
\begin{document}

\title{Bipolar polaron pair recombination in P3HT/PCBM solar cells}

\author{Alexander J. Kupijai}
\email{alexander.kupijai@wsi.tum.de}
\author{Konstantin M. Behringer}%
\affiliation{Walter Schottky Institut and Physik-Department, Technische Universität München, Am Coulombwall 4, 85748 Garching, Germany}%

\author{Michael Corazza}
\author{Suren A. Gevorgyan}
\author{Frederik C. Krebs}
\affiliation{Department of Energy Conversion and Storage, Technical University of Denmark, Frederiksborgvej 399, 4000 Roskilde, Denmark}%

\author{Martin Stutzmann}%
\author{Martin S. Brandt}%
\email{brandt@wsi.tum.de}
\affiliation{Walter Schottky Institut and Physik-Department, Technische Universität München, Am Coulombwall 4, 85748 Garching, Germany}%

\date{\today}

\begin{abstract}
The unique properties of organic semiconductors make them versatile base materials for many applications ranging from light emitting diodes to transistors. The low spin-orbit coupling typical for carbon-based materials and the resulting long spin lifetimes give rise to a large influence of the electron spin on charge transport which can be exploited in spintronic devices or to improve solar cell efficiencies. Magnetic resonance techniques are particularly helpful to elucidate the microscopic structure of paramagnetic states in semiconductors as well as the transport processes they are involved in. However, in organic devices the nature of the dominant spin-dependent processes is still subject to considerable debate. Using multi-frequency pulsed electrically detected magnetic resonance (pEDMR), we show that the spin-dependent response of P3HT/PCBM solar cells at low temperatures is governed by bipolar polaron pair recombination involving the positive and negative polarons in P3HT and PCBM, respectively, thus excluding a unipolar bipolaron formation as the main contribution to the spin-dependent charge transfer in this temperature regime. Moreover the polaron-polaron coupling strength and the recombination times of polaron pairs with parallel and antiparallel spins are determined. Our results demonstrate that the pEDMR pulse sequences recently developed for inorganic semiconductor devices can very successfully be transferred to the study of spin and charge transport in organic semiconductors, in particular when the different polarons can be distinguished spectrally.\end{abstract}

\maketitle

\begin{figure*}[ht]
\centering
\includegraphics[width=\linewidth]{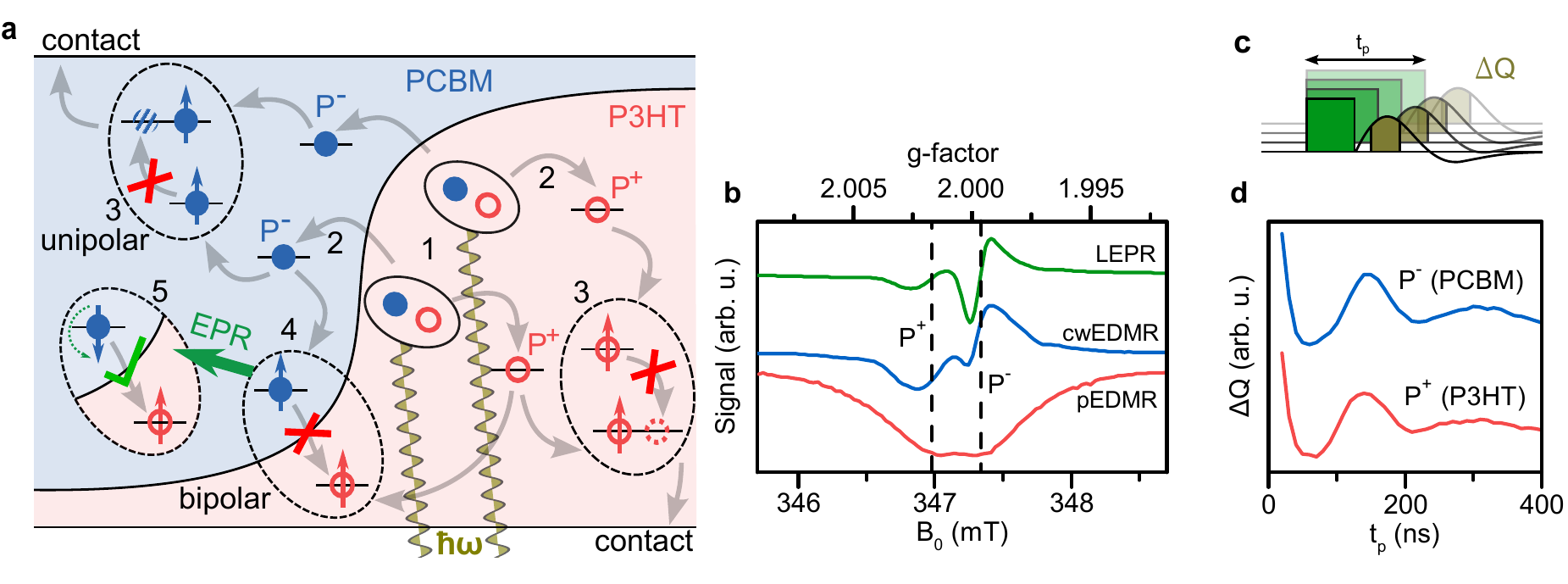}
\caption{\textbf{Spin-dependent processes in P3HT/PCBM heterostructures.} \textbf{a}, Schematic diagram of charge transport in P3HT/PCBM bulk heterojunction devices. After absorption of photons (1), the excitons generated diffuse to the P3HT/PCBM interface, where the charges are separated, eventually forming polarons P$^+$ in P3HT and P$^-$ in PCBM (2). Spin-dependent processes involving these polarons can either be unipolar, where polarons of the same charge form a bipolaron (3), or bipolar, where two polarons of different charge annihilate (4). The Pauli principle demands that both hopping and recombination are only possible if the spins of the two polarons are antiparallel, polarons with parallel spin orientation will not be allowed to undergo the hopping or recombination process. As sketched exemplarily for the recombination, the respective process can be enhanced by magnetic resonance, when the spin of either polaron is flipped and the parallel spin configuration is changed to an antiparallel one (5). \textbf{b}, Light-induced electron paramagnetic resonance (LEPR), cw electrically detected magnetic resonance (cwEDMR) and pulsed EDMR (pEDMR) spectra of P3HT/PCBM. All spectra exhibit the characteristic resonance peaks at the g-values of the positive and the negative polarons in P3HT and PCBM, respectively (depicted by vertical dashed lines). \textbf{c}, Pulse sequence for the electrical detection of the coherent driving of spins (Rabi oscillations). The current transients following the resonant manipulation of the spin system are integrated, yielding a charge $\Delta Q$ as EDMR signal. \textbf{d}, Electrically detected Rabi oscillations as a function of the microwave pulse length $t_p$ used to drive the magnetic resonance of the negative or positive polaron.}
\label{spin}
\end{figure*}

In addition to their wide application as light emitting diodes (LEDs)~\cite{reineke_white_2009}, organic semiconductors also are promising materials for solar cells due to their mechanical flexibility and their potential low cost production~\cite{sondergaard_roll--roll_2012}. One important basic physical aspect is their spin-orbit coupling which is low compared to e.g.~III-IV compound semiconductors. The resulting long spin lifetimes lead to a pronounced spin dependence of charge transfer processes, providing the opportunity for further improvements of solar cell efficiency~\cite{xu_photovoltaic_2008,gonzalez_improved_2015} or the development of spintronic devices~\cite{Yoo2010}. In both organic molecules and polymers, the lack of rigidity leads to a significant structural relaxation when excess charges are present, such as the formation of positively or negatively charged polarons. Magnetic resonance techniques so far are limited to time scales of $\gtrsim\SI{1}{\micro\second}$, so that they are particularly helpful to understand the physics of polarons in organic semiconductors. Two different fundamental spin-dependent processes involving polarons can occur, as sketched in Fig.~\ref{spin}a: recombination of a bipolar polaron pair~\cite{frankevich_polaron-pair_1992,Schulten_magnetic_1976} and hopping, where a doubly charged polaron is created from a unipolar polaron pair~\cite{bobbert_bipolaron_2007,behrends_bipolaron_2010}. While the former process gives rise to spin-dependent changes in the charge or rather the polaron carrier concentration, the latter results in a spin-dependent change of polaron mobility. Which of the two polaronic  processes is observed in specific organic diodes, both LEDs and solar cells, is currently subject to considerable debate~\cite{boehme_challenges_2013}. The distinction is particularly difficult in the case where the spectroscopic signatures of positive and negative polarons are very similar, as in the case of PPV~\cite{boehme_pulsed_2009,behrends_bipolaron_2010,baker_differentiation_2011,tedlla_understanding_2014}. A more direct assignment should be possible in organic devices where the two polarons can be distinguished spectroscopically. This also allows the direct transfer of pulse sequences developed for the study of charge transport and recombination in inorganic semiconductors such as amorphous and crystalline Si~\cite{boehme_theory_2003,stegner_electrical_2006,hoehne_spin-dependent_2010,suckert_electrically_2013}. Here  we study polaron transport and kinetics in P3HT/PCBM solar cell structures, fabricated both by spin coating and printing, using pulsed electrically detected magnetic resonance (pEDMR), identify bipolar polaron pair formation as the dominant spin-dependent process at low temperature and determine recombination and coherence times.

The negative and positive polarons P$^-$ and P$^+$ in PCBM and P3HT exhibit g-factors of 1.9996 and 2.0017, respectively, as observed by light-induced electron paramagnetic resonance (LEPR)~\cite{krinichnyi_light-induced_2009,dietmueller_light-induced_2009} and transient EPR (trEPR)~\cite{behrends_direct_2012,niklas_photoinduced_2015}. For typical linewidths of $\SI{0.9}{\milli\tesla}$, these g-factors allow a near-perfect spectroscopic separation at X-band frequencies around $\SI{9}{\giga\hertz}$, as desired for our study. Figure~\ref{spin}b shows a comparison of the signatures of the two polarons as observed by us as a function of the static magnetic field $B_0$ with (i) LEPR at $\SI{50}{\kelvin}$, where the magnetization of the sample is measured, with (ii) continuous wave (cw) EDMR at $\SI{10}{\kelvin}$, where the DC photocurrent through the sample is detected, and with (iii) pulsed EDMR also at $\SI{10}{\kelvin}$, where the photocurrent transient after a microwave pulse is investigated. In all cases illumination was performed with the red light of a LED. LEPR and cwEDMR are performed with the help of magnetic field modulation, generating first derivative spectra. As expected, in all these experiments the two different polarons are clearly and distinctly observed.

\begin{figure*}[ht]
\centering
\includegraphics[width=\linewidth]{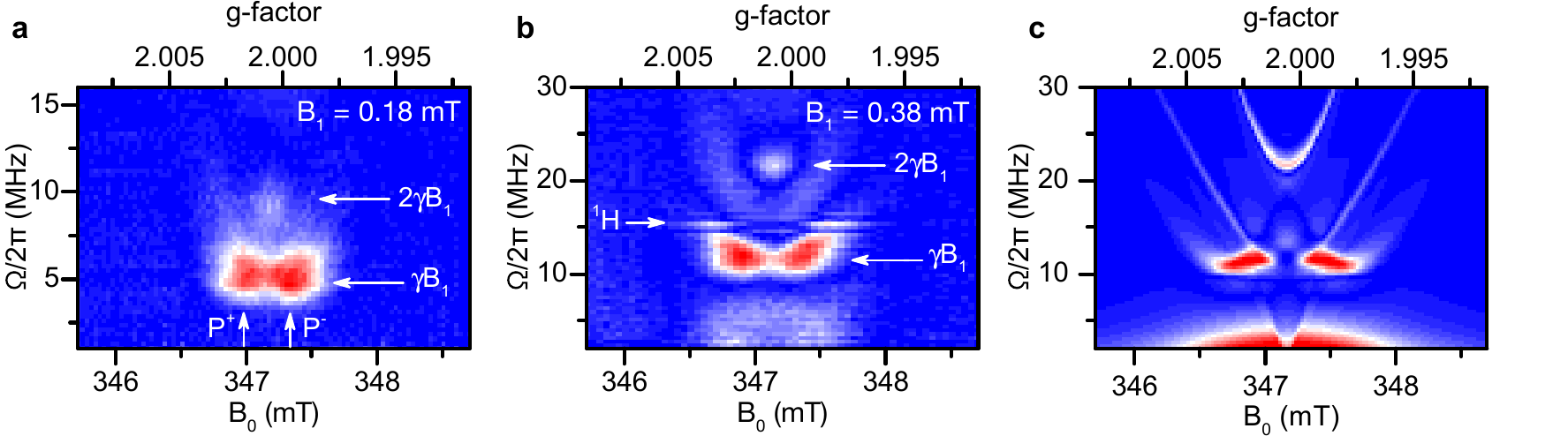}
\caption{\textbf{Spin-locking in P3HT/PCBM.} \textbf{a}, Fast Fourier transforms of Rabi oscillation measured by pEDMR for different magnetic fields $B_0$ at a low microwave power. Two strong signals attributed to P$^+$ and P$^-$ are observed at the fundamental Rabi frequency $\Omega=\gamma B_1$. Already a weak spin-locking signal with $\Omega=2\gamma B_1$ is visible at a central magnetic field. \textbf{b}, At higher microwave power, the spin-locking signal becomes more prominent. In addition, a nuclear magnetic resonance signal with a frequency of $\sim \SI{14.8}{\mega\hertz}$ emerges, caused by hyperfine interaction with $^1$H protons in the organic material (see supplementary information). \textbf{c}, Simulation of the Rabi map in Fig.~\ref{elsor}b, using a superoperator Liouville space formalism and assuming an exchange coupling between the positive and negative polarons of $J/2\pi=\SI{2}{\mega\hertz}$.}
\label{elsor}
\end{figure*}

\section*{Spin locking}
Pulsed magnetic resonance experiments can provide a wealth of further information in addition to the spectroscopic identification of the polarons. Figure~\ref{spin}d shows the results of a Rabi oscillation experiment in pEDMR with the pulse sequence sketched in Fig.~\ref{spin}c, where the length $t_p$ of the microwave pulse is varied. On both, the positive and the negative polaron, virtually identical oscillations with a period of $\sim\SI{130}{\nano\second}$ are observed. By repeating these experiments with different microwave powers or at different $B_0$, we can distinguish between uni- and bipolar pair formation in P3HT/PCBM structures based on so-called spin locking. Spin locking occurs if the spectral excitation width is large enough to flip both spins belonging to a spin pair at the same time~\cite{mccamey_hyperfine-field-mediated_2010,behrends_bipolaron_2010}. This is the case when the spectral width (i) becomes comparable to the inhomogeneous linewidth of the resonance in the case of unipolar pair formation or (ii) becomes so large that both resonances can be excited in the opposite case of bipolar pair formation. Since in EDMR the singlet symmetry of the polaronic spin pair is measured~\cite{kaplan_explanation_1978}, the simultaneous driving of magnetic resonance on both constituents of the spin pair leads to an effective doubling of the Rabi oscillation, so that its frequency is $\Omega=2\gamma B_1$, where $\gamma$ is the gyromagnetic ratio and $B_1$ the microwave field. For weak $B_1$, however, only one of the spins will be driven, so that in this case we expect the classic $\Omega=\gamma B_1$. This limit also enables the determination of $B_1$, which scales to higher microwave powers $P$ as $B_1\propto\sqrt{P}$.

The distinction between uni- and bipolar pair formation can now be made easily if the two polaron resonances are spectrally separated: In the case of hopping between polarons of identical charge, the spin-locking signal at high $B_1$ will appear at the same spectral position as that of the polarons, while the corresponding signal for recombination involving polarons of two different charges will be observed at magnetic fields between the resonance positions of the positively and negatively charged polarons. This is indeed observed in P3HT/PCBM; Figures~\ref{elsor}a and b show Fourier transforms of Rabi oscillation experiments for different magnetic fields at two different microwave powers (below, we will refer to this type of plot as a Rabi map). Already for low $B_1$ (Fig.~\ref{elsor}a) a small trace of spin locking is visible in between the resonances of the polarons in P3HT and PCBM as expected for bipolar pairs. For higher microwave intensities (Fig.~\ref{elsor}b) this spin-locking peak becomes even more prominent.

This assignment is further supported by simulation. Following the approach of Limes {\it et al.}, the spin Hamiltonian is solved numerically using the superoperator Liouville space formalism~\cite{limes_numerical_2013}. Assuming an exchange interaction $J/2\pi$ of $\SI{2}{\mega\hertz}$ between the positive and the negative polaron, Larmor frequencies corresponding to the g-factors given above and a $B_1$ of $\SI{0.38}{\milli\tesla}$, the Rabi frequency map shown in Fig.~\ref{elsor}c is obtained. In addition, a Gaussian distribution of Larmor frequency differences of $\SI{25}{\mega\hertz}$ is used to account for the inhomogeneous linewidths of $\sim \SI{0.9}{\milli\tesla}$ of both polarons at X-band frequencies. The simulation reproduces the experimental Rabi maps well, in particular with respect to the central bipolar spin-locking feature at $\Omega/2\pi\approx\SI{22}{\mega\hertz}$. The parabola-like wings visible in both experiment and simulation are caused by the fact that Rabi oscillations speed up off resonance~\cite{stegner_electrical_2006}. The exact lineshapes of the polarons, variations in their coupling as well as details of the microwave pulse shape and the effects of the resonator are not taken into account in the simulation, likely causing the remaining differences between simulation and experiment.

In order to compare the pEDMR results for differently processed samples, these Rabi measurements were repeated on roll-to-roll produced P3HT/PCBM solar cells~\cite{sondergaard_roll--roll_2013}. These cells show essentially the same spectra and Rabi oscillation behavior (data in supplementary information). Therefore, the specific production process seems to have a negligible influence on the results of our experiments, allowing to conclude that bipolar recombination is the dominant spin-dependent process in P3HT/PCBM structures at low temperatures. The only significant difference between spin-coated and roll-to-roll printed devices is the lower pEDMR signal intensity of the latter, notable in the lower signal-to-noise ratio realizable in the experiments, an indication that less recombination takes place in these optimized devices.

\begin{figure*}[ht]
\centering
\includegraphics[width=0.277777\linewidth]{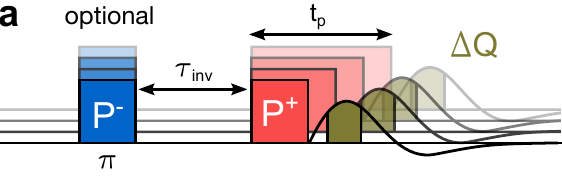}
\hspace{0.032\linewidth}
\includegraphics[width=0.277777\linewidth]{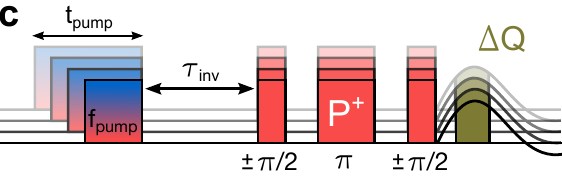}
\hspace{0.032\linewidth}
\includegraphics[width=0.277777\linewidth]{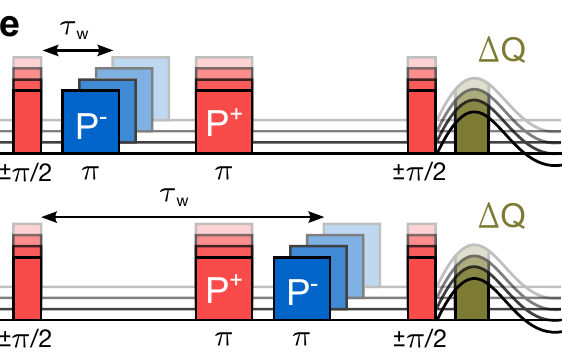}

\hspace{0.032\linewidth}

\includegraphics[width=\linewidth]{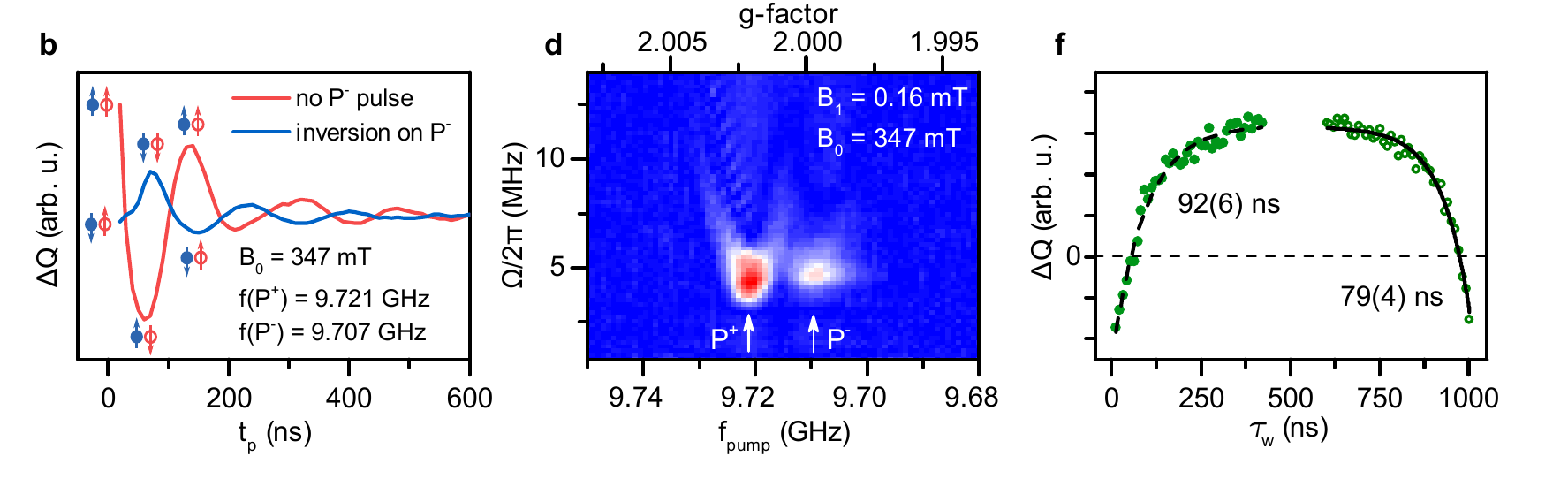}
\caption{\textbf{Double resonance experiments using ELDOR and DEER.} \textbf{a}, Pulse sequence for electron double resonance (ELDOR) using two microwave frequencies to address the positive and negative polarons for a certain magnetic field $B_0$ separately. \textbf{b}, Without a leading pulse on P$^-$, a Rabi oscillation experiment on P$^+$ leads to the same oscillation already observed in Fig.~\ref{spin}d. With a leading $\pi$ pulse inverting the P$^-$ spin, the oscillation is inverted, as expected for bipolar polaron pair recombination. \textbf{c}, Modification of the ELDOR pulse sequence for frequency mapping. A pulse of length $t_\mathrm{pump}$ and frequency $f_\mathrm{pump}$ is followed by an echo sequence on the P$^+$ resonance to measure the influence of the first pulse on the positive polaron. \textbf{d}, Corresponding ELDOR Rabi map. A clear and spectrally well resolved Rabi oscillation driving the P$^-$ resonance is observed on the P$^+$. The waiting time $\tau_\mathrm{inv}$ between the pulses with different frequency was $\SI{200}{\nano\second}$ in both b and d. \textbf{e}, Pulse scheme for the determination of the spin coupling strength via double electron electron resonance (DEER). \textbf{f}, The echo intensity as a function of the delay between the initial $\pi/2$ pulse on P$^+$ and the $\pi$ inversion pulse on P$^-$. The time dependence allows to estimate the coupling strength to $\sim\SI{2}{\mega\hertz}$.}
\label{eldor}
\end{figure*}

\begin{figure*}[ht]
\centering
\includegraphics[width=0.277777\linewidth]{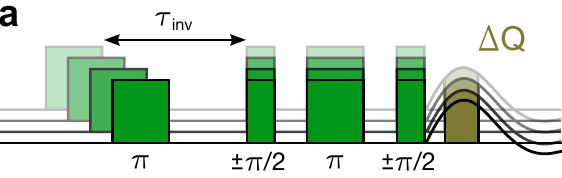}
\hspace{0.032\linewidth}
\includegraphics[width=0.277777\linewidth]{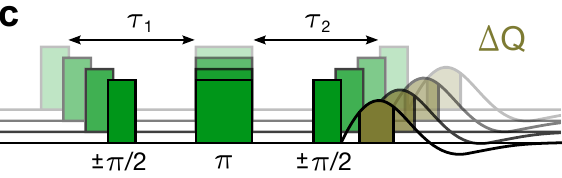}
\hspace{0.095\linewidth}
\includegraphics[width=12.6cm]{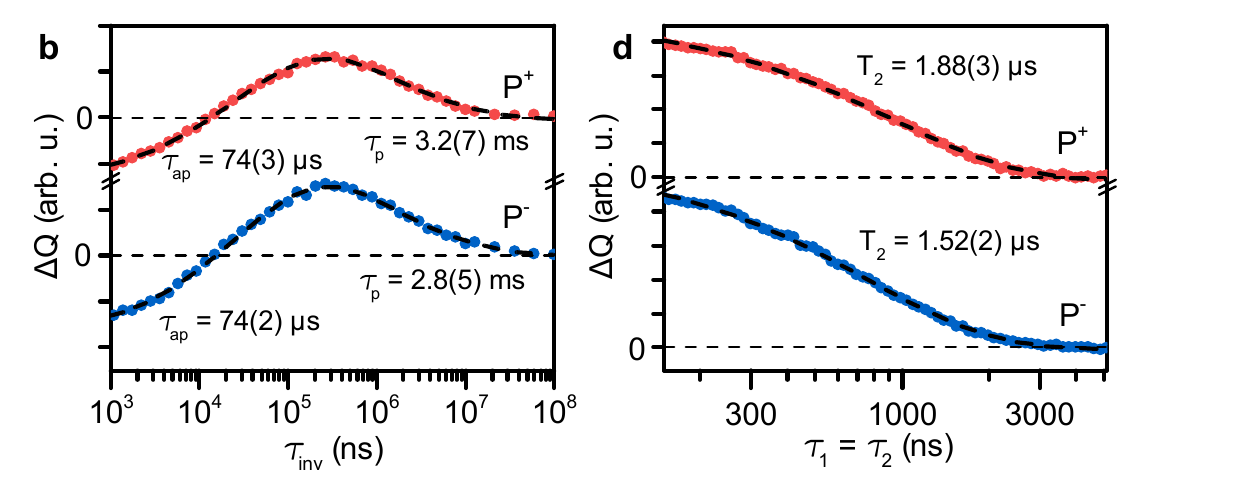}
\caption{\textbf{Polaron kinetics.} \textbf{a}, Pulse sequence used to determine the antiparallel and the parallel recombination times $\tau_\mathrm{ap}$ and $\tau_\mathrm{p}$, respectively, of P$^+$P$^-$ polaron pairs, based on an inversion recovery experiment. \textbf{b}, Typical time constants of $\SI{74}{\micro\second}$ for $\tau_\mathrm{ap}$ and $\SI{3}{\milli\second}$ for $\tau_\mathrm{p}$ are found experimentally. \textbf{c}, Standard echo sequence including a final $\pi/2$ projection pulse for the measurement of the coherence time $T_2$. \textbf{d}, Both polarons exhibit similar decoherence properties with $T_2\approx\SI{1.7}{\micro\second}$.}
\label{times}
\end{figure*}

\section*{Addressing polarons separately}
Electron double resonance (ELDOR)~\cite{hoehne_spin-dependent_2010} and double electron electron resonance (DEER)~\cite{suckert_electrically_2013} experiments using pulse sequences, where the recombination partners are addressed separately by different microwave frequencies, show even more convincingly that bipolar polaron recombination is observed. The results for the spin-coated cells obtained with ELDOR are summarized in Fig.~\ref{eldor}a and b. The experiments were performed at a fixed $B_0$, with one of the microwave frequencies tuned to resonantly excite the positive polaron and the other the negative counterpart. First, a Rabi oscillation experiment is performed as in Fig.~\ref{spin}c, here on the P$^+$, plotted in Fig.~\ref{eldor}b. The pEDMR signal initially decreases from a level, indicative of the spins in the polaron pair being parallel to each other, to a level corresponding to antiparallel spins due to a $\pi$ inversion pulse on one of the spins, the P$^+$. If, preceding this Rabi experiment, a $\pi$ pulse is applied to the other spin, the P$^-$ in this case, the Rabi oscillation will start in an antiparallel pair configuration, turning to a parallel configuration after a pulse of length $\pi$ for the P$^+$ Rabi experiment. The result will be an inversion of the Rabi oscillation, which is indeed observed in Fig.~\ref{eldor}b. If, on the other hand, unipolar hopping would be observed, the spin symmetry of P$^+$P$^+$ pairs observed in the Rabi oscillation would remain the same irrespective of changes of the spin state of a P$^-$, which is not observed. The degree of inversion in the bipolar case depends on the degree to which the P$^-$ ensemble is addressed by the corresponding microwave pulse, which is about $\SI{70}{\percent}$ in our case.

These ELDOR experiments can also be performed as a function of the microwave frequencies used. The pulse sequence of such an ELDOR Rabi map is shown in Fig.~\ref{eldor}c. With a frequency $f_\mathrm{pump}$ and for varying times $t_\mathrm{pump}$ a microwave pulse is applied to the P3HT/PCBM cell, followed by an echo sequence on the P$^+$ resonance used to measure the P$^+$ spin state~\cite{hoehne_time_2013}. With respect to the pulse sequence in Fig.~\ref{eldor}a, this is effectively only a time-inverted experiment, but allows to use phase cycling and lock-in detection~\cite{hoehne_lock-detection_2012}, very significantly improving the signal-to-noise ratio. The corresponding ELDOR Rabi map is shown in Fig.~\ref{eldor}d. As in the case of the single frequency Rabi map, a signal is seen in the Fourier transform corresponding to $\Omega=\gamma B_1$. The signal is strong when $f_\mathrm{pump}$ also excites P$^+$ and, in agreement with Fig.~\ref{eldor}b, somewhat weaker when exciting P$^-$. Most importantly, there is a clear signal minimum between the two resonances, even more so as in Fig.~\ref{elsor}a showing the corresponding single-frequency Rabi map. Spurious excitation of P$^+$ in the experiments of Fig.~\ref{eldor}a and b due to the small spectral overlap of P$^+$ and P$^-$ can therefore be excluded.

While the ELDOR experiments discussed show that indeed bipolar spin pairs are formed, DEER experiments are able to quantitatively estimate the strength of the coupling between the two polarons. The corresponding pulse sequence is shown in Fig.~\ref{eldor}e, where an inversion pulse on P$^-$ is applied either in the first or the second evolution period of an echo sequence on P$^+$. The change in the magnetic environment of the P$^+$ during its evolution period due to the change of the coupled spin state of P$^-$ will lead to an inversion of the signal depending on the point of time, when the P$^-$ spin is flipped. While ideally a well defined spin-spin interaction manifests itself in an oscillatory behavior of the echo signal as a function of the shift $t_w$ (see Fig.~\ref{eldor}e), a broad distribution of coupling strengths typically leads to an exponential decay of the echo intensity, still allowing to estimate a characteristic coupling from the decay constant. The results of the DEER experiments on the P3HT/PCBM cells are shown in Fig.~\ref{eldor}f, with characteristic decay times of 92 and 79~ns in the case of the inversion pulse in the first and second evolution times, respectively, corresponding to a typical coupling strength of $\SI{2}{\mega\hertz}$ between the negative and positive polaron. This value was already used in the simulation of Fig.~\ref{elsor}c. The lack of an oscillation in the DEER results indicates a broad distribution of coupling strengths, which was not taken into account in the simulation and is one of the reasons for the remaining differences between experiment and simulation in Fig.~\ref{elsor}.

Nevertheless, both types of dual frequency experiments, ELDOR and DEER, performed here clearly support the initial argument based on the single frequency spin-locking measurements that spin-dependent processes in bipolar polaron pairs are observed. Since cwEDMR experiments without lock-in amplification (data not shown) show a reduction of the photocurrent under resonance condition, we can attribute the resonances observed to bipolar polaron pair recombination, since hopping would lead to a resonant increase of the mobility and therefore the conductivity.

\section*{Polaron recombination kinetics}
The identification that the dominant spin-dependent process in the P3HT/PCBM solar cells at low temperatures is the bipolar polaron pair recombination allows to transfer a wealth of pulse sequences for the study of spin and recombination kinetics, which were mostly developed to investigate the recombination involving donors and defects in crystalline Si~\cite{paik_T1_2010,hoehne_time_2013}, also to organic semiconductors. In particular, inversion recovery experiments can be used to determine the recombination times of polaron pairs with antiparallel and parallel spin orientation $\tau_\mathrm{ap}$ and $\tau_\mathrm{p}$, respectively (Fig.~\ref{times}a). The steady state spin population during illumination, which, due to the fast spin-allowed recombination of antiparallel spin pairs, mostly consists of parallel spin pairs, is first inverted into antiparallel pairs by a $\pi$ inversion pulse. The recombination of those pairs is then measured with an echo sequence. As expected for pair recombination, the corresponding experiments performed on the positive and the negative polaron agree, with $\tau_\mathrm{ap}\approx\SI{74}{\micro\second}$ (Fig.~\ref{times}b). Since again the excitation width is not large enough to invert the whole ensemble, parallel spin pairs persist, whose recombination time $\tau_\mathrm{p}\approx\SI{3}{\milli\second}$ can therefore be determined in the same experiment. Due to the variation in the coupling strengths, the recombination kinetics are best described by stretched exponentials exp$(-(t/\tau)^n)$~\cite{hoehne_time_2013}. In our experiment, we observe $n_\mathrm{ap}=0.45$ and $n_\mathrm{p}=0.25$ for the recombination of antiparallel and parallel spins, respectively.

The echo sequence can also be used to measure the decoherence time of the polaron spins (Fig.~\ref{times}c). We find values of $T_2=\SI{1.9}{\micro\second}$ for the positive polaron in P3HT and $T_2=\SI{1.5}{\micro\second}$ for the negative counterpart in PCBM (Fig.~\ref{times}d), obtained by fitting with normal exponential functions.

\section*{Conclusions}
We have shown that pulsed EDMR can be used to distinguish between bipolar polaron pair recombination and unipolar bipolaron hopping transport in organic semiconductors when the resonant signatures of the differently charged polarons can be separated spectroscopically. The assignment to bipolar recombination is made on the basis of different single- and multi-frequency experiments including spin locking, ELDOR and DEER. In addition to the identification of the dominant spin-dependent process, the advanced pulse sequences used allow the determination of the strength of the coupling between the polarons, the lifetimes of polaron pairs with antiparallel and parallel spin orientations as well as the coherence time of the spins. The pEDMR experiments performed here are highly versatile and were applied to both laboratory scale devices as well as cells fabricated on an industrial scale, demonstrating the generality of the spin-dependent process identified. This method is neither limited to the P3HT/PCBM material system nor the $\SI{10}{\kelvin}$ temperature range investigated here. In particular, optical excitation with ns pulses might allow to study the temporal evolution of the polaron coupling, effectively monitoring the formation of the polaron pairs, or even the investigation of charge transfer levels with pEDMR methods. Optical excitation resonant with the charge transfer levels will be helpful in this respect~\cite{street_photoconductivity_2011}. Therefore, the present work can be an important stepping stone for further investigations of related material systems to understand charge transport processes as well as the microscopic steps involved in photovoltaic energy conversion and lighting applications in organic devices.

\section*{Methods}
The study presented in this work was performed on bulk heterojunction organic solar cells consisting of poly(3-hexylthiophen-2,5-diyl) (P3HT) and phenyl-C$_{61}$-butyric acid methyl ester (PCBM). The spin-coated solar cells were produced on lithographically structured indium tin oxide (ITO)-on-glass substrates with the dimension of approx.~$4\times\SI{40}{\milli\meter\squared}$ in order to fulfill geometric requirements of the spectrometer. A hole transport layer of  poly(3,4-ethylenedioxythiophene) polystyrene sulfonate (PEDOT:PSS) was spin coated on the substrate first. The spin coating of the active blend (P3HT:PCBM = 1:1) at 720~rpm, which results in a thickness of around $\SI{100}{\nano\meter}$, was performed under inert gas conditions in an argon glovebox. Finally, a top electrode of approx.~$\SI{120}{\nano\meter}$ aluminum was evaporated inside the glovebox. The active area of the cell is about $\SI{4}{\milli\meter\squared}$, the rest of the substrate is needed to electrically contact the cell without disturbing the microwave resonator. Without further encapsulation the samples were bonded onto a sample holder and transferred into the spectrometer, exposing the sample to air in darkness for less than 5 minutes. In the spectrometer, the helium atmosphere of the cryostat prevents degradation.

The printed reference samples were produced on a flexible 130 micron thick polyester substrate (Melinex ST506 from Dupont-Teijin). The solar cell stack was prepared in an inverted architecture using full roll-to-roll processing of all layers at high speed following a method similar to literature reports~\cite{krebs_25th_2014,krebs_freely_2013}. The device stack comprises six printed layers starting with a flexographically printed slanted silver comb front electrode structure~\cite{krebs_25th_2014}, rotary screen printed semitransparent PEDOT:PSS, slot-die coated zinc oxide, slot-die coated P3HT:PCBM (1:1), rotary screen printed PEDOT:PSS and finally a rotary screen printed oppositely slanted silver comb back electrode structure. The solar cells, which are produced on a much larger area than needed, were cut into smaller pieces to fit into the spectrometer.

The samples for the LEPR measurement were blends of P3HT/PCBM (1:1) drop-casted on glass, thus no working solar cells. These samples were sealed in a glass tube under inert gas.

The pEDMR measurements were performed at $\SI{10}{\kelvin}$ and under a negative bias voltage of $\SI{-3}{\volt}$ in a custom-built pEDMR spectrometer based on a commercial pulsed X-band resonator. At this temperature, all solar cells tested behaved as a photoconductor. Microwaves from two separate sources were shaped to rectangular pulses using mixers and a multi-channel pulse generator and then amplified by a traveling wave tube amplifier. Photocurrent transients were amplified by trans\-impedance and low noise voltage amplifiers, band pass filtered and recorded with a digitizer card. Boxcar integration of the transients (typically from $\SI{1.5}{\micro\second}$ to $\SI{15}{\micro\second}$) generated a charge $\Delta Q$, the primary pEDMR signal. The samples were illuminated by continuous red LED light throughout all magnetic resonance experiments, with the exception of the inversion recovery experiment, where the light was switched off $\SI{50}{\micro\second}$ before the start of the microwave pulse sequence, in order to avoid photogeneration of new charge carriers during the recombination time measurements. The evolution time for the echo sequences was $\SI{300}{\nano\second}$, except for the DEER experiment, where it was $\SI{500}{\nano\second}$. The shot repetition frequency of the pulsed experiments was $\SI{1}{\kilo\hertz}$, again with exception of the inversion recovery experiment, which was repeated with $\SI{10}{\hertz}$. Four cycle phase cycling was used in all experiments based on echo sequences, cycling the phase of the leading and trailing $\pi /2$ pulses by $\SI{180}{\degree}$.

The LESR measurements were performed at $\SI{50}{\kelvin}$ in a commercial EPR spectrometer. The cwEDMR measurements were performed at $\SI{10}{\kelvin}$ in the same spectrometer.

\section*{Acknowledgements}
This work was supported by the Eurotech Alliance through the International Graduate School of Science and Engineering (IGSSE), project Interface Science for Photovoltaics (ISPV). The authors thank Willi Aigner, David Franke, Florian Hrubesch, Andreas Sperlich, Max Suckert and Stefan Väth for helpful discussions.

\bibliography{bipolar_literature}

\pagebreak
\widetext
\section*{Supplementary information}

\begin{figure*}[ht]
\centering
\includegraphics[width=\linewidth]{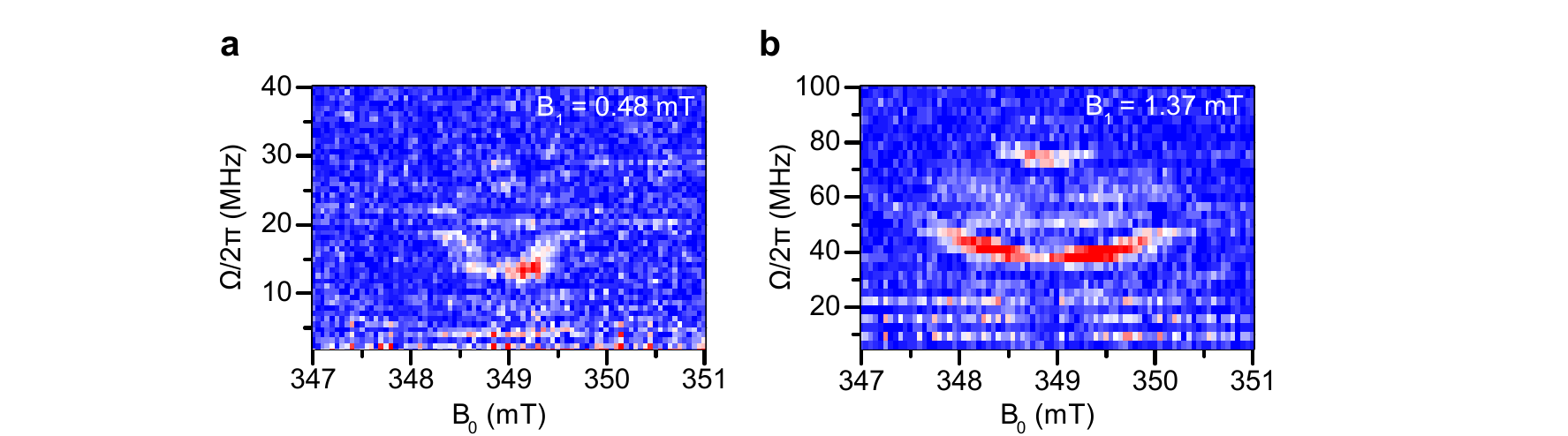}
\caption{\textbf{Spin locking in printed P3HT/PCBM solar cells.} \textbf{a-b}, Fast Fourier transforms of Rabi oscillation measurements by pEDMR for different magnetic fields. Again, the two signals corresponding to P$^+$ and P$^-$ are observed, with the central spin-locking signal appearing at higher microwave powers.}
\label{dtu}
\end{figure*}

The printed organic solar cells show similar Rabi oscillation maps compared to the spin-coated solar cells, demonstrating that bipolar polaron pair recombination is the dominant spin-dependent process also in P3HT/PCBM blends fabricated with a technique different from spin coating. The signal-to-noise ratio is worse than in spin-coated cells, most likely due to a better optimization of the printed cells. Higher $B_1$ fields were needed to obtain sufficient signal intensity.

\begin{figure*}[ht]
\centering
\includegraphics[width=9cm]{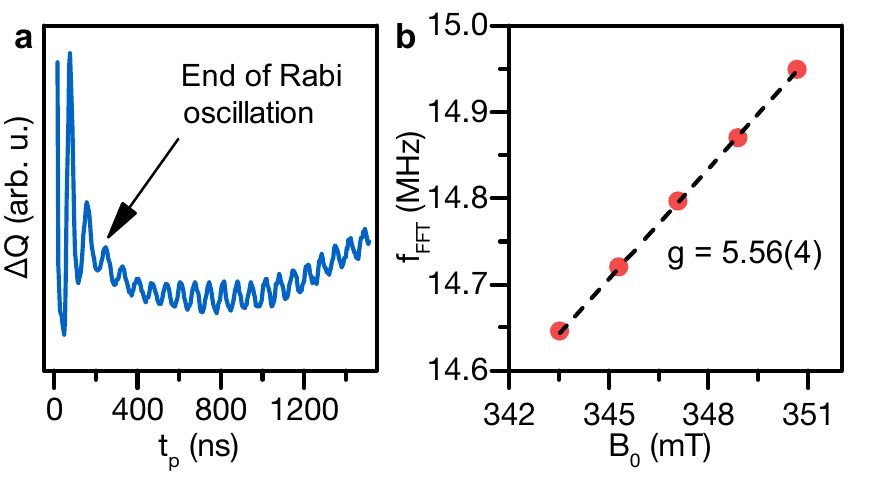}
\caption{\textbf{Nuclear magnetic resonance of hydrogen.} \textbf{a}, pEDMR signal as a function of the pulse length $t_p$. The strong initial Rabi oscillation dephases, leaving an oscillation persisting to long $t_p$. \textbf{b}, The frequency of the persisting oscillation $f_\mathrm{FFT}$ depends linearly on the magnetic field $B_0$, with a proportionality characteristic for the nuclear magnetic resonance of protons.}
\label{nmr}
\end{figure*}

The additional feature in the fast Fourier transform spectra in Fig.~\ref{elsor}b at $\Omega/2\pi\approx\SI{14.8}{\mega\hertz}$ can be attributed to hyperfine interaction with hydrogen nuclear spins. This interaction gives rise to additional oscillations which persist even after the Rabi oscillations have dephased (Fig.~\ref{nmr}a). These oscillations do not change their period with the microwave $B_1$ field, but a clear dependence of the frequency on the static magnetic field $B_0$ is visible (Fig. \ref{nmr}b). This dependence allows the attribution to a nuclear magnetic resonance with a g-factor of 5.56(4), which is in very good agreement with the proton g-factor~\cite{mohr_codata_2012} and previous findings on PPV~\cite{malissa_room-temperature_2014}.

\end{document}